\documentclass[sigconf]{acmart}

\setcopyright{none} % No copyright notice required for submissions
\acmConference[]{}{}{}
\acmYear{2026}

\usepackage[british]{babel}
\usepackage[T1]{fontenc}
\usepackage[utf8x]{inputenc}
\usepackage{graphicx}
\usepackage{listings, xcolor}
\usepackage{fix-cm}\rmfamily  % [tex.stackexchange.com/q/334368]
\usepackage{paralist}
\usepackage{graphics}
\usepackage{amsmath}
\usepackage{booktabs}
\usepackage{siunitx}
\usepackage{tikz}
\usetikzlibrary{patterns}
\usepackage{multirow}
\usepackage[ruled,linesnumbered,noend]{algorithm2e}

\usepackage{lipsum}
\usepackage{gensymb}
\usepackage{colortbl}
\usepackage{tabularx}
\usepackage{pdflscape}
\usepackage{subcaption}
\usepackage{array}
\usepackage{enumitem}
\usepackage{listings}
\usepackage{ragged2e}

\usepackage{tikz}
\usetikzlibrary{shapes, arrows, positioning}

\tikzstyle{block} = [rectangle, draw, text width=0.95\columnwidth, align=left, minimum height=1cm]
\tikzstyle{arrow} = [thick,->,>=stealth]

\usepackage{pgfplots} 
\pgfplotsset{
	compat=1.9,
	compat/bar nodes=1.8,
}
\usepackage{threeparttable}
\usepackage{pgfplotstable}
\usetikzlibrary{patterns}
\usepackage{xcolor}
\usepackage[dvipsnames,table]{xcolor} 
\usepackage{amsfonts}
\usepackage{pifont}
\usepackage{xspace}
\usepackage{orcidlink}
\usepackage{todonotes}
\usepackage{comment}
\usepackage[multiple]{footmisc}
\usepackage{bigfoot}
\usepackage{rotating}
\usepackage{multicol}
\usepackage{caption}

\newcommand\goalstatement{\textit{The goal of this study is to aid software practitioners in understanding the reliability of dependency adoption trust signals, such as download counts and contributor activity, by conducting a multivocal review of 252 Google Search sources and 870 Reddit threads.}}

\newcounter{rqcounter}
\newcounter{subrqcounter}[rqcounter]
\newcommand{\newrq}[2]{%
    \begin{description}[leftmargin=2.8em]
    \refstepcounter{rqcounter}%
    \setcounter{subrqcounter}{0}%
    \item[\textbf{RQ\arabic{rqcounter}:}] {\em #2}\label{#1}
    \end{description}
}
\newcommand\qt[1]{\textcolor{blue}{\textit{``#1''}}}
\newcommand\doc[1]{{\footnotesize$\langle$#1$\rangle$}}
\newcommand\signal[1]{\textcolor{OliveGreen}{\emph{#1}}}

\newcounter{findingcounter}

\usepackage{titlesec}
\titleformat{\subsection}{\normalfont\large\bfseries}{\thesubsection}{.5em}{}

\newcommand{\newfinding}[2]{%
    \refstepcounter{findingcounter}%
    \label{#1}\vspace{.4em}%
    \textbf{\textcolor{Purple}{Finding \#\arabic{findingcounter}:}} \textcolor{Purple!95}{#2}%
}

\begin{document}
\title{The Software Supply Chain as a Market for Lemons: A Multivocal Review of Trust Signal Collapse} % TODO: replace with your title

\author{Ranindya Paramitha}
\affiliation{%
  \institution{North Carolina State University}
  \city{Raleigh, NC}
  \country{USA}}
\email{rparami@ncsu.edu}

\author{Christian K\"astner}
\affiliation{%
  \institution{Carnegie Mellon University}
  \city{Pittsburgh, PA}
  \country{USA}}
\email{kaestner@cs.cmu.edu}

\author{Laurie Williams}
\affiliation{%
  \institution{North Carolina State University}
  \city{Raleigh, NC}
  \country{USA}}
\email{lawilli3@ncsu.edu}

\begin{abstract}
%Background
% Practitioners evaluating open-source dependencies may rely on cheap trust signals, e.g., stars, download counts, contributor activity, and documentation quality, in place of direct code inspection. 
% This common practice assumes that those signals reflect the genuine trustworthiness of dependencies in the software supply chain.
Practitioners evaluating open-source dependencies rely on cheap trust signals, e.g., stars, download counts, and contributor activity, as substitutes for direct code inspection, assuming those signals reflect genuine trustworthiness.
Prior work has documented individual signal gaming, but the landscape of collapses across all dependency-adoption signals, as well as the ecosystem's response, remains unexplored.
% Goal
\goalstatement\xspace
% Method
% We coded the Google Search sources for gaming mechanisms, adversarial intent, and ecosystem response across eight signal categories.
% We also ran LLM-assisted coding for both corpora.
% Results
After coding the corpora, we find that cheap trust signals collapse under three simultaneous forces: adversarial manipulation, gaming techniques indistinguishable from legitimate behavior, and non-adversarial AI-driven inflation.
% (the latter two account for 52.7\% of documented mechanisms).
The documented responses are more advice than actual action: 54.6\% of Google Search sources contain advice on what practitioners should do, with no actual action taken.
Responses proposed substituting one cheap signal for another or aggregating multiple signals, which are now also gameable.
Non-adversarial inflation, i.e., degradation caused by the emergence of legitimate AI tooling, lacks documented actual behavior change in either corpus. 
% Takeaway
The gap between known remedy and actual practice points toward a market for lemons: when faking signals costs less than earning them, good and bad dependencies become indistinguishable.
Relying on individual practitioners to verify the cheap signals is not sustainable.
Costlier signals, such as cryptographic attestation, should be made mandatory so that they become the default for all, not a voluntary choice for the few.\looseness=-1
% This condition calls for institutional enforcement of costlier signals, such as cryptographic attestation, rather than continued reliance on individual practitioners to verify the cheap, gameable signals.
\end{abstract}

% TODO: replace this section with code generated by the tool at https://dl.acm.org/ccs.cfm
% \begin{CCSXML}
% <ccs2012>
%    <concept>
%        <concept_id>10011007.10011074.10011134.10003559</concept_id>
%        <concept_desc>Software and its engineering~Open source model</concept_desc>
%        <concept_significance>500</concept_significance>
%        </concept>
%    <concept>
%        <concept_id>10002978.10003022</concept_id>
%        <concept_desc>Security and privacy~Software and application security</concept_desc>
%        <concept_significance>500</concept_significance>
%        </concept>
%    <concept>
% <concept_id>10003456.10003462</concept_id>
% <concept_desc>Social and professional topics~Computing / technology policy</concept_desc>
% <concept_significance>300</concept_significance>
% </concept>
% </ccs2012>
% \end{CCSXML}

% \ccsdesc[500]{Software and its engineering~Open source model}
% \ccsdesc[500]{Security and privacy~Software and application security}
% \ccsdesc[300]{Social and professional topics~Computing / technology policy}

 % TODO: replace with your keywords

\maketitle

\begin{figure}[t]
    \centering
    \includegraphics[width=\columnwidth]{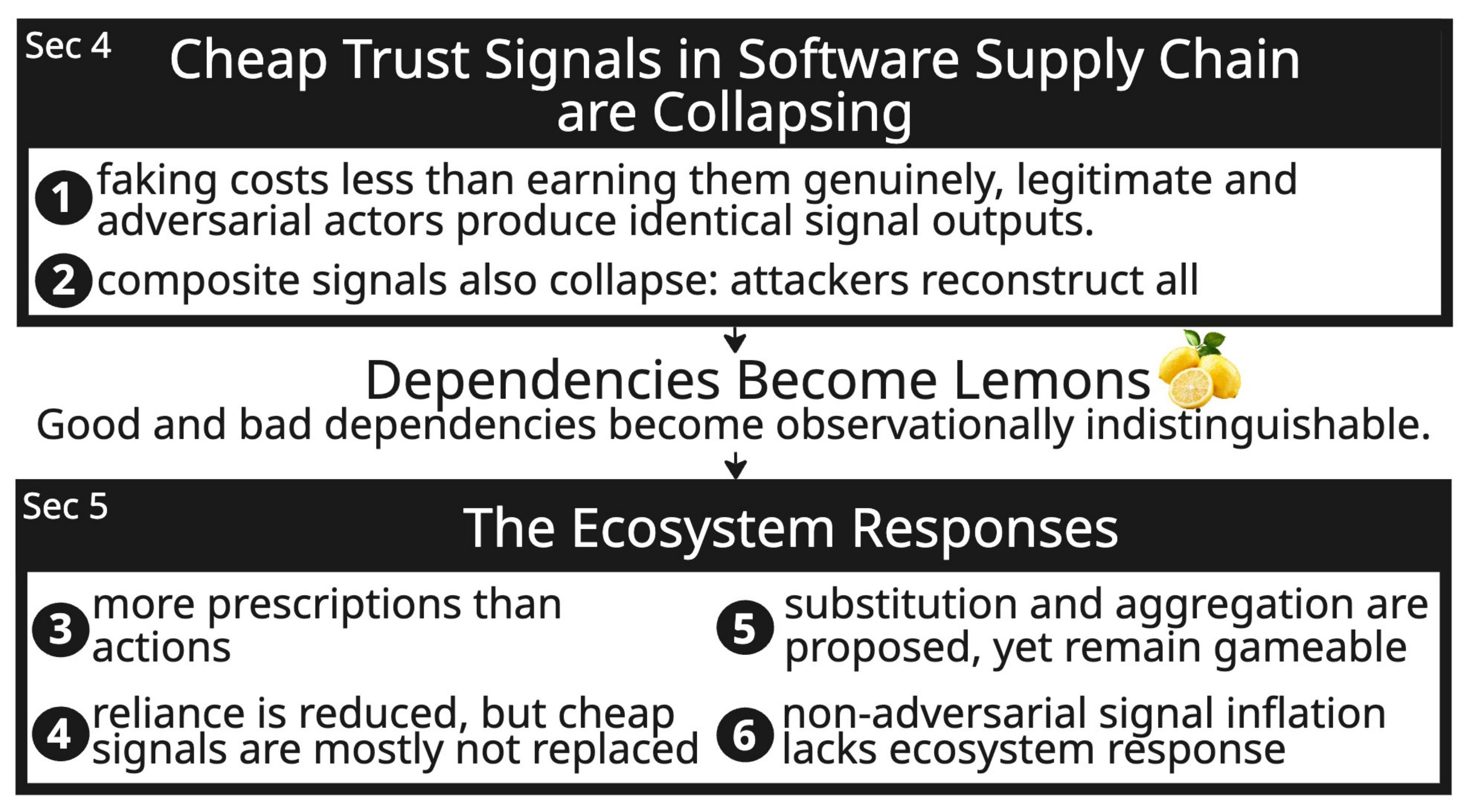}
    \begin{flushleft}
    \scriptsize\textit{Image credit:} Lemon image by
    \textcolor{blue}{\href{https://www.vecteezy.com}{Vecteezy}}.
    \end{flushleft}
    \caption{Overview: Cheap Signals Degrade, Dependencies Become Lemons, Ecosystem Responds.}
    \label{fig:summary}
\end{figure}

\section{Introduction}
\label{sec:introduction}
When adopting an open-source dependency, do you read the source code, or do you rely on signals such as stars and contributor activity to infer its trustworthiness?
Prior empirical work confirms that practitioners do the latter~\cite{pashchenko2020qualitative,wermke2022committed,hamer2025trusting}, and until recently, that was a reasonable choice.
In May 2026, a rogue AI agent operating under a Fedora developer's compromised credentials spent weeks submitting patches and responding to reviewer objections with LLM-generated justifications that ``eventually overwhelmed the maintainer into merging the fix''~\cite{lwn2026fedora}.
% The motive remains unconfirmed to this day: the same behavior, i.e., a new contributor slowly gaining trust, submitting plausible changes, responding persistently to objections, could equally be a negligently deployed coding tool, a hijacked account, or a deliberate XZ Utils-style attack in its preparatory phase~\cite{freund2024xz}.
% The unsettling part is not that the attack succeeded, but that the behavior was indistinguishable from that of a legitimate contributor, regardless of intent.
The motive remains unconfirmed because the behavior was indistinguishable from that of a legitimate contributor, regardless of intent, similar to the xz utils incident~\cite{freund2024xz}.
This incident is not isolated; it is a symptom of how the software supply chain is structured.
What is distinctive about the current moment is not only that trust signals for adopting dependencies are being gamed (which has always happened), but also that AI is eroding so many signals so fast, even all at once.
With AI, producing convincing stars, commit histories, documentation, and community endorsements at scale is now cheap. 
The cost gap between honest and malicious signal production is collapsing, bringing the open source dependencies ecosystem towards a market for lemons, where developers are unable to distinguish trustworthy packages from untrustworthy ones~\cite{akerlof1970market}.
Prior work has documented the manipulation of individual signals: fake 
stars~\cite{he2026six}, inflated downloads~\cite{tenable2026downloadpumping}, 
injected documentation~\cite{kaspersky2025gitvenom}, and maintainer 
social engineering~\cite{freund2024xz}.
Yet, a systematic consolidation across the full range of dependency adoption signals and of the ecosystem’s responses has not been done. 
\goalstatement\xspace
% Our Multivocal Literature Review (MLR) systematically combines formal academic publications with informal 
% practitioner sources~\cite{garousi2019guidelines} to capture knowledge documented outside academic venues.
% We first selected eight \signal{trust signals} that prior empirical work has identified as proxies developers consult when evaluating dependencies: \signal{GitHub stars}, \signal{download counts}, \signal{documentation quality}, \signal{contributor activity} (both AI-inflated and adversarially mimicked), \signal{platform-awarded achievement badges}, \signal{issue response quality}, and \signal{community endorsement}.
% After developing search queries, we manually retrieved 252 sources via Google Search, spanning peer-reviewed 
% publications, vendor reports, blog posts, and forum threads.
% We also separately retrieved 870 relevant Reddit threads from 95,190 candidates using LLM-assisted relevance filtering.
Our MLR~\cite{garousi2019guidelines} combines formal academic publications with practitioner sources, across eight trust signals: \signal{GitHub stars}, \signal{download counts}, \signal{documentation quality}, \signal{AI-inflated contributor activity}, \signal{adversarially mimicked contributor activity}, \signal{platform-awarded achievement badges}, \signal{issue response quality}, and \signal{community endorsement}.
We coded the Google Search sources using a deductive two-pass scheme addressing two research questions:
% Our Multivocal Literature Review (MLR)~\cite{garousi2019guidelines} addressed two research questions.
% \begin{itemize}
%     \item \textbf{RQ1:} What mechanisms are documented as causing cheap 
%     trust signals in the software supply chain to collapse?
%     \item \textbf{RQ2:} What does the multivocal literature record as the 
%     ecosystem's response to this collapse, and where does that response 
%     fall short?
% \end{itemize}

\newrq{rq1}{What mechanisms are documented as causing cheap trust signals in the software supply chain to collapse?}
% {As suggested by previous work, we find that the economics of cheap trust signals have inverted: faking now costs less than earning.
% Legitimate AI tools, deliberate attackers, and honest actors can all produce identical signal outputs, making collapse a structural property of the signals themselves rather than purely a detection problem.
% Sophisticated attacks compound this by reconstructing the entire trust envelope simultaneously, rendering composite evaluation ineffective.}

\newrq{rq2}{What does the multivocal literature record as the ecosystem's response to this collapse, and where does that response fall short?}
% {
% The documented response is overwhelmingly prescriptive: 54.6\% of Google Search sources provide recommendations on what practitioners should do but describe no practitioners actually doing so.
% The recommended substitutes are predominantly cheap signals already documented as gameable, and signal inflation caused by legitimate AI tooling lacks documented responses that lead to actual behavior change.}
% The proposed substitutes are predominantly cheap signals already documented as gameable in this same corpus, and non-adversarial signal inflation lacks documented response with actual behavior change.}

The coding captured the manipulation technique, adversarial intent, ecosystem response type, and changes in reliance on signals for each source.
Intercoder reliability was assessed by comparing the first-author manual codings with LLM codings on a 40-source sample across three calibration rounds, achieving over 70\% agreement across all RQ1 fields.
The 252 Google Search sources and the 870 Reddit threads were also each independently analyzed using LLM-assisted thematic coding~\cite{qiao2025thematic}.
This analysis produces separate codebooks of codes and practitioner quotes that provide qualitative support and direct practitioner voice for the quantitative patterns identified through the manual coding scheme.

Our analysis confirmed previous works that the economics of cheap trust signals have inverted: faking now costs less than earning.
Legitimate AI tools, deliberate attackers, and honest actors can all produce identical signal outputs, making collapse a structural property of the signals themselves rather than purely a detection problem.
Sophisticated attacks compound this by reconstructing the entire trust envelope simultaneously, rendering composite evaluation ineffective.
The documented response is predominantly advice without action: 54.6\% of Google Search sources provide recommendations on what practitioners should do but describe no practitioners actually doing so.
The recommended substitutes are predominantly cheap signals already documented as gameable, and signal inflation caused by legitimate AI tooling lacks documented responses that lead to actual behavior change.
% The proposed substitutes are predominantly cheap signals already documented as gameable in this same corpus, and non-adversarial signal inflation lacks documented response with actual behavior change.
% After this introduction section (\S\ref{sec:introduction}), some background and related works are discussed in Section~\ref{sec:background_relwork}. 
% Section~\ref{sec:methodology} discussed the methodology of this work, including the corpus construction and coding/analysis methods. 
% The results of the analysis are then elaborated in the result sections (\S\ref{sec:results-rq1}--\ref{sec:results-rq2}).
% We acknowledge the limitations of the study in Section~\ref{sec:limit} and conclude the paper by discussing the results and implications in Section~\ref{sec:discussion}.
% The paper is concluded with a conclusion and future work section (\S\ref{sec:conclusion}).
% An ethical consideration note is also included at the end of the paper.

\noindent In summary, our contributions are:
\begin{itemize}[leftmargin=*]
\item A multivocal corpus of 252 Google Search sources and 870 Reddit threads on trust signal manipulation and ecosystem response across eight dependency adoption signals, with an available codebook and annotated dataset;
\item Corpus-based findings on the mechanisms causing cheap trust signals to collapse, documenting ten manipulation mechanism categories across eight signals and distinguishing adversarial, ambiguous, and non-adversarial degradation vectors;
\item Corpus-based findings on the ecosystem's documented responses to trust signal collapse, characterizing response types, trust direction changes, and proposed signal alternatives across eight signals;
\item Role-based recommendations for practitioners, platform operators, tool builders, and policy makers, grounded in signaling theory~\cite{spence1973}, Goodhart's Law~\cite{goodhart1975problems}, and Akerlof's market for lemons~\cite{akerlof1970market}.\looseness=-1
\end{itemize}

We discuss background and related works in Section~\ref{sec:background_relwork} and methodology in Section~\ref{sec:methodology}. The results of the analysis are then elaborated in the result sections (\S\ref{sec:results-rq1}--\ref{sec:results-rq2}).
We acknowledge the limitations of the study in Section~\ref{sec:limit} and conclude the paper with a discussion and implications in Section~\ref{sec:discussion}.

\section{Background and Related Works}
\label{sec:background_relwork}

\subsection{Trust and Signals in the Software Supply Chain}
According to Mayer et al.~\cite{mayer1995integrative}, trust is the willingness of one party to accept vulnerability to the actions of another, based on positive expectations about that party's conduct. 
In the software supply chain, open-source software reuse scales through dependency: developers extend trust by incorporating code written by others, relying on the assumption that the dependency behaves as expected~\cite{parnas1972criteria}.
Modern applications routinely incorporate hundreds of transitive dependencies~\cite{soto2021comprehensive}, making direct verification impractical and creating a structural reliance on observable \signal{signals} as proxies for trustworthiness before adoption.
Signaling theory~\cite{spence1973} explains why these signals can be informative: a signal carries information only when the marginal cost of producing it is lower for high-quality producers than for low-quality ones, also known as the single-crossing condition.
When this condition holds, a separating equilibrium exists in which trustworthy and untrustworthy dependencies are distinguishable, allowing downstream users to infer quality without direct inspection~\cite{spence1973}.
Prior empirical work has documented which signals practitioners actually consult when evaluating dependencies.
GitHub stars and download counts are used as indicators of popularity and adoption~\cite{pashchenko2020qualitative,qiu2019signals,wermke2022committed}. 
Others also use contributor activity, reputation, and maintenance cadence as indicators of project health~\cite{pashchenko2020qualitative, qiu2019signals,wermke2022committed, dabbish2012social, hamer2025trusting}.
Qiu et al.~\cite{qiu2019signals} also report that practitioners check documentation (README) quality before adoption.
Pashchenko et al.~\cite{pashchenko2020qualitative} show that practitioners also consult community platforms such as Stack Overflow and Reddit to obtain endorsements and recommendations for dependencies.

When the low-quality producers can produce the signal at a cost that no longer separates them from the high-quality ones, the separating equilibrium collapses into a pooling equilibrium in which the signal carries no meaningful information~\cite{spence1973}.
The market drifts toward the lemon equilibrium Akerlof predicted: downstream users cannot distinguish good dependencies from bad, and the ecosystem as a whole loses the trust that made cheap reuse possible~\cite{akerlof1970market}.
Goodhart's Law~\cite{goodhart1975problems} describes the mechanism by which this collapse is triggered: once practitioners rely on a metric (signal) to make decisions, it becomes worth gaming and thus ceases to measure what it was supposed to measure.
Together, three theoretical lenses: signaling costs, Goodhart's targeting dynamic, and adverse selection, ground this study's analysis and distinguish it from prior empirical work on how practitioners behave when trusting open-source dependencies~\cite{wermke2022committed,pashchenko2020qualitative,sammak2023developers}.

\subsection{Signal Gaming in the Software Supply Chain}
Prior work has documented the manipulation of individual trust signals in isolation.
Coordinated bot networks inflated star counts~\cite{he2026six}.
Automated install loops and mirror infrastructure inflated download figures~\cite{sonatype2026sscr}.
AI-generated README files simulated project legitimacy~\cite{kaspersky2025gitvenom}.
Multi-year social engineering campaigns could earn co-maintainer status~\cite{freund2024xz}.
Planted User-Generated Content (UGC) text could manipulate AI retrieval pipelines~\cite{zhang2026warp}.
Each of these studies examines a single signal, mechanism, or incident.
The contribution of this study is a cross-signal synthesis that documents the landscape of manipulation mechanisms, the intentions behind those mechanisms, and the ecosystem's response, as recorded in the literature.

\subsection{Ecosystem Responses of Signal Manipulation}
The empirical studies documenting signal use~\cite{pashchenko2020qualitative,wermke2022committed,sammak2023developers,hamer2025trusting} have not examined what happens when practitioners learn that those signals could be gamed. These prior works document current practices on adopting dependencies, not a response to the disclosed failure of a signal.
Existing attack taxonomies~\cite{ohm2020backstabber,ladisa2023sok} focus on artifact-level compromises (code injection, credential theft, and build-infrastructure attacks) and their corresponding technical safeguards. 
Trust signal manipulation, which targets the perception of project trustworthiness rather than the integrity of the artifact itself, falls outside their scope.
These two bodies of literature have not been connected.
Some platform and regulatory responses to signal manipulation are visible in the literature, such as platform decisions about which signals to display~\cite{pypidownloads}.
Yet, they have been documented as individual case studies rather than in the full landscape of manipulation mechanisms.
Whether these responses could collectively address the collapse or a specific signal manipulation mechanism, and how widely they are adopted/advised, remain open questions that this study addresses.

\subsection{Multivocal Reviews in Software Engineering}

A Multivocal Literature Review (MLR) systematically includes both formal academic publications (white literature) and informal practitioner sources such as blog posts, forum threads, and vendor reports~\cite{garousi2019guidelines}.
MLRs are particularly well-suited to topics in which practitioner knowledge is documented primarily outside academic venues. This condition applies directly to trust signal manipulation, where pricing data for fake-star services, incident reports, and platform/practitioners' responses appear predominantly in grey literature.
Prior MLRs in software engineering have applied the methodology to topics including DevSecOps practices~\cite{zhao2024identifying, myrbakken2017devsecops}, non-technical debt~\cite{saeeda2024multivocal}, and software testing~\cite{garousi2016and}.
These MLRs establish grey literature as a valid and necessary source for capturing practitioner knowledge in software engineering.
In the software supply chain security domain, existing studies, including attack taxonomies/systematic literature reviews~\cite{ladisa2023sok,reichert2024software} and research direction surveys~\cite{williamsDirection2025}, draw primarily from academic databases. 
Trust signal manipulation and practitioner responses to it, which are documented predominantly in grey literature, have not been a focus of these reviews.
This study applies the MLR methodology to the software supply chain trust-signal domain, combining 252 Google Search sources and 870 Reddit threads across eight signal categories to provide a cross-signal, cross-tier synthesis.

\section{Methodology}
\label{sec:methodology}
We conducted a Multivocal Literature Review (MLR), which systematically includes both formal academic publications (white literature) and informal practitioner sources such as blog posts, forum threads, public videos, vendor reports, and grey literature~\cite{garousi2019guidelines}.
We chose an MLR over a traditional systematic literature review to capture practitioners' responses, which are documented predominantly outside academic venues.
For example, pricing data for fake-star services, practitioner accounts of signal gaming, and platform responses would appear more in grey literature sources than in white literature sources.

\subsection{Corpus Construction}
To construct a corpus for our MLR, we first define the signals in scope, then perform the source search and retrieval using Google Search. We then applied inclusion and exclusion criteria to get the final corpus. We also performed an additional gray literature retrieval on Reddit using a script adapted from a publicly available tool~\cite{grayliteraturetool}.

\noindent \textbf{Signals in Scope.}
We organized the corpus around eight trust signals that were identified through a preliminary review of the software engineering literature on dependency adoption decisions~\cite{dabbish2012social,qiu2019signals,pashchenko2020qualitative,wermke2022committed,hamer2025trusting}.
% We organized the corpus around eight trust signals that prior empirical research has identified as proxies developers use when evaluating open-source dependencies for adoption~\cite{dabbish2012social,qiu2019signals,pashchenko2020qualitative,wermke2022committed,hamer2025trusting}, 
The signals are summarized in Table~\ref{tab:signals}.
The separation between \signal{S3a} and \signal{S3b} reflects a coding decision made during querying: sources documenting AI-driven activity inflation (non-adversarial) and sources documenting deliberate activity spoofing (adversarial) required separate treatment because their mechanisms and implications differ.

\noindent \textbf{Search and Retrieval.}
For each of the eight signals, we identified candidate search vocabulary through an LLM-assisted exploratory pre-search (with \texttt{claude-sonnet-4-6}), following a consistent structure: signal name combined with a manipulation or fakery term and a platform or context qualifier (e.g., \textit{GitHub stars manipulation inflate popularity} for \signal{S1a:stars}). 
Some signals required multiple iterations before supply-chain-context vocabulary emerged.
% The resulting terms were logged for traceability; each seed query in the main retrieval phase traces to a term observed in pre-search results rather than assumed in advance.
% We first constructed pre-search queries combining the signal name with manipulation-related terms (e.g., \textit{fake GitHub stars}, \textit{npm download inflation} and response-related terms (e.g., \textit{choosing dependency based on X}).
Using that vocabulary, we constructed a maximum of three each of (1) anchor queries, i.e., narrow queries that name a specific known incident/paper to find a specific source and snowball (backward/forward) from that source; and (2) seed queries, i.e., broad, single-concept queries to build a corpus across grey literature.
We executed each anchor and seed query on Google Search and took all links on Google's first page, including academic papers, blog posts, social media discussions, and YouTube videos.
We supplemented the corpus by (1) targeted browsing of practitioner platforms, including Hacker News and dev.to, Medium, and Stack Overflow; (2) snowballing from the anchor sources.
We also retrieved vendor and institutional reports on open-source and software supply chain security from Sonatype~\cite{sonatype2026sscr} and Snyk~\cite{snyk2024oss}.
Retrieval was conducted between July 1st and 19th, 2026, yielding 321 candidate sources.
All queries are reported in the supplementary materials~\cite{zenodo}.

\noindent \textbf{Inclusion and Exclusion.} 
Sources were included if they met all of the following criteria:
(1) the source discusses at least one of the eight trust signals by name or by clear functional description; and (2) the source is publicly accessible in English.
Sources were excluded if they discussed software quality or security in general terms without reference to any of the eight signals.
After applying inclusion and exclusion criteria, 252 sources were retained for coding.
We also applied a post hoc software supply chain relevance classification using an LLM-assisted classifier (\texttt{claude-sonnet-4-6}) to assess whether each source is related to the software supply chain (e.g., dependency adoption).
This classification was used for corpus characterization only and did not affect inclusion decisions: sources documenting signal collapse in non-software supply chain contexts were retained because they provide the only available evidence for signals, particularly \signal{S3a:activity-legitimate}, \signal{S4:platform-awarded-achievements}, and \signal{S5:issue-response}.
The manipulation of those signals has not yet been widely discussed in the formal literature as a software supply chain concern.

\noindent \textbf{Additional Gray Literature.}
We also separately retrieved 95,190 Reddit threads from 2023-2026, of which 870 were judged by an LLM as relevant to this study with high confidence and had a post score $>$ 0.
This retrieval is using a script adapted from a publicly available tool for gray literature review~\cite{grayliteraturetool}.

\begin{table}[]
    \centering
    \footnotesize
    \caption{Signals in Scope and Retrieved Google Search Sources}
    \begin{tabular}{p{0.3\columnwidth}|p{0.53\columnwidth}|r} \hline
        Code & Longer explanation & \# \\ \hline
        \signal{S1a:stars} & GitHub stars as popularity signal~\cite{dabbish2012social,qiu2019signals,wermke2022committed} & 40 \\
        \rowcolor{gray!20}
        \signal{S1b:downloads} & Download counts as popularity signal~\cite{pashchenko2020qualitative} & 46 \\
        \signal{S2:documenta\-tion} & documentation quality (README, API docs, in-repository examples)~\cite{qiu2019signals} & 14 \\
        \rowcolor{gray!20}
        \signal{S3a:activity-legitimate} & contributor activities (commit frequency and history)~\cite{wermke2022committed,hamer2025trusting,
pashchenko2020qualitative}, specifically legitimate activity that can be inflated because of AI tools usage & 33 \\
        \signal{S3b:activity-mimicry} & contributor activities (commit frequency and history)~\cite{wermke2022committed,hamer2025trusting}, specifically mimicry activity that inflates the numbers &  54 \\
        \rowcolor{gray!20}
        \signal{S4:platform-awarded-achievements} & achievement badges & 11 \\
        \signal{S5:issue-response} & issue response time and quality~\cite{pashchenko2020qualitative,wermke2022committed} & 17 \\
        \rowcolor{gray!20}
        \signal{S6:community-endorsement} & Stack Overflow, Reddit, forum recommendations~\cite{pashchenko2020qualitative} & 37\\ 
        \hline
        \multicolumn{2}{l}{Total} & 252 \\ \hline
    \end{tabular}
    \label{tab:signals}
\end{table}

\noindent \textbf{Final Corpus.}
After collection and filtering, we have a final Google Search corpus comprising 252 sources across 8 signals. \signal{S3b:activity-mimicry} is the largest category (54 sources), reflecting the volume of incident reporting around commit spoofing and social engineering. On the other hand, \signal{S2:documentation} is the smallest (14 sources), a gap we acknowledge as a corpus limitation.
Source tiers follow the MLR convention~\cite{garousi2019guidelines}: Tier 1 comprises peer-reviewed academic publications (n=49; 19.4\%), Tier 2 comprises grey literature from identifiable organizations with editorial oversight (vendor reports, institutional publications, and journalistic sources; n=63; 25.0\%), and Tier 3 comprises practitioner-authored grey literature without formal editorial review (blog posts, forum threads, and social media posts; n=137; 54.4\%).
The post hoc relevance classification found that 152 sources (60.3\%) are directly related to software supply chain security, while 100 (39.7\%) document the relevant signals in adjacent contexts without an explicit software supply chain framing.
Additionally, we have 870 relevant Reddit threads from the gray literature review tool.
Both corpora are made available in the replication package~\cite{zenodo}.
In the result, we refer to sources from Google Search as \doc{DX} and from the tool's Reddit threads as \doc{RX}, with X as the ID number.

\subsection{Analysis}
We first developed a coding scheme and then coded the 252 Google Search sources. We also run a thematic analysis coding using a tool inspired by~\cite{qiao2025thematic} on both the 252 Google Search sources and 870 Reddit threads separately.

\noindent \textbf{Google Search Corpus Coding Scheme.}
We developed deductively a two-pass coding scheme addressing each research question separately.
RQ1 coding captures the mechanism by which each signal is manipulated or structurally degraded; RQ2 coding captures the actors' responses in the ecosystem. 
The coding scheme is summarized in Table~\ref{tab:coding-scheme}.
For RQ1, each source was first coded in the field
\textit{rq1\_applies} (YES / PARTIAL / NO), which indicates whether the source documents a specific manipulation mechanism or only references signal unreliability in passing.
For sources with \textit{rq1\_applies} YES / PARTIAL, we also code (1) \textit{mechanism\_tech}: primary/secondary manipulation technique; and (2) \textit{mechanism\_adversarial}: intent behind the mechanism.
For RQ2, each source was first coded with \textit{rq2\_applies} (YES/NO), indicating whether the source contains a practitioner, platform, or researcher voice in the response.
For the sources with \textit{rq2\_applies} YES, we also coded (1)
\textit{statement\_type}: the type of the documented response, distinguishing advice from reported enacted actions); (2) \textit{trust\_direction}: the documented change in signal reliance; and (3) \textit{alt\_signal}: free-text specific alternative signal/s proposed in the source.

\begin{table*}[]
    \centering
    \footnotesize
    \caption{Coding Scheme for the Google Search Corpus}
    \begin{tabular}{l|p{.82\linewidth}}
        \hline
        Code & Explanation \\ \hline
        \multicolumn{2}{l}{\textbf{RQ1 — Mechanism Technique (MT): manipulation method/s}} \\ \hline
        MT-BOT         & Coordinated fake-account networks or automated non-AI scripts that generate stars, downloads, or activity signals at scale \\
        MT-AI-AGENT    & AI coding agents (e.g., Claude Code, Cursor) inflating commits, pull requests, or activity signals as a structural side-effect of normal development \\
        MT-SPOOF       & Falsification of commit metadata, authorship attribution, or timestamps to impersonate reputable contributors\\
        MT-CRED        & Credential or token theft enabling attackers to publish malicious package versions under a trusted maintainer identity\\
        MT-SOC-ENG     & Social engineering to acquire trust in a signal or actor, e.g., gaining trust as a maintainer or gaining trust for a specific package\\
        MT-UGC-PLANT   & Planting promotional text in user-generated content platforms (Reddit, Stack Overflow, forums) for both people and AI retrieval \\
        MT-DOC-FAKE    & AI-generated or injected README files, documentation, or in-repository content simulating project legitimacy \\
        MT-PKG-INJ     & Injection of malicious code into an existing package via a compromised release, typosquatting, or namespace confusion \\
        MT-BADGE-GAME  & Activity to unlock GitHub achievement badges or other platform-awarded trust markers \\
        MT-HALLUC      & Exploitation of LLM-hallucinated package names: registering non-existent packages that AI coding tools recommend\\
        MT-OTHER       & Documented manipulation mechanism not captured by the above categories\\ \hline
        \multicolumn{2}{l}{\textbf{RQ1 — Mechanism Adversarial (MA): intent behind the mechanism}} \\ \hline
        MA-ADV         & Deliberate deception with intent to mislead: the mechanism is specifically deployed to harm or deceive receivers \\
        MA-AMB         & Mechanism technique can be used for both legitimate and adversarial intent at the signal level: honest and dishonest actors produce identical outputs \\
        MA-NON         & Non-adversarial structural inflation: signal degradation as a side-effect of legitimate tooling with no deceptive intent \\ \hline \hline
        \multicolumn{2}{l}{\textbf{RQ2 — Statement Type (ST): epistemic status of the documented response}} \\ \hline
        A — Stated only  & Advice, recommendation, or stated concern without any described enacted action; includes researcher advice, vendor guidance, and practitioner opinion\\
        B — Revealed only & A specific action described by the source author without explicitly stated concern \\
        C — Coupled      & A described action explicitly linked to a stated concern by the source author \\
        D — Analyst-inferred  & The coder infers a response from indirect evidence; the source does not directly state a response \\ \hline
        \multicolumn{2}{l}{\textbf{RQ2 — Trust Direction (TD): recorded change in signal reliance}} \\ \hline
        no change           & Source acknowledges no change in signal use \\
        signal abandoned    & Source records ceasing use of the signal with no stated replacement \\
        signal downweighted & Source records continued use of the signal with reduced weight or increased scrutiny, e.g., added verification or other signals as composite \\
        signal reaffirmed   & Source argues the signal remains useful or reliable despite documented manipulation \\
        signal replaced     & Source records switching to a different signal entirely, no longer consulting the original\\ \hline
    \end{tabular}
    \label{tab:coding-scheme}
\end{table*}

\noindent \textbf{Google Search Corpus Coding Procedure} proceeded in two phases. \textbf{1. Two coders coded a selected 40 sources.} We pre-selected a sample of 40 sources, with a balanced number of sources per signal and document tier. These 40 sources are coded both manually by the first author and by a large language model (LLM) (\texttt{claude-sonnet-4-6}) via the Anthropic API for RQ1 and RQ2. The LLM was prompted with the full codebook and the extracted source text. We assessed intercoder reliability in this sample by comparing the first author's manual codings with LLM codings. We calibrate the LLM prompt over three rounds, and for the fields with agreement $>70\%$ (all RQ1 coding fields), we use the LLM as a single coder in the next phase. \textbf{2. Single coding on the rest.} In this phase, LLM coded the rest of the corpus for RQ1 (agreement on sample $>70\%$), and the first author coded the rest for RQ2.

We report Krippendorff's alpha and Cohen's kappa (both at the nominal level) alongside percentage agreement, using percentage agreement as the primary threshold criterion: fields were accepted for single-coding when agreement exceeded 70\%, consistent with recommendations for skewed-distribution coding tasks where kappa is artificially depressed by base-rate effects~\cite{krippendorff1980,cohen1960coefficient}.
All RQ1 closed-code fields exceeded the 70\% threshold after three rounds of prompt calibration: \textit{rq1\_applies} (82.5\%), \textit{mechanism\_tech\_1} (82.5\%), \textit{mechanism\_tech\_2} (70.0\%), and \textit{mechanism\_adversarial} (80.0\%).
RQ2 fields were coded manually by the first author, as the reliability of the LLM for \textit{trust\_direction} was insufficient (agreement $<70\%$).

\noindent \textbf{LLM-assisted Thematic Coding.}
For both the Google Search corpus and the 870 Reddit threads, we ran the LLM-assisted thematic coding tool (inspired by and built on Qiao et al.'s~\cite{qiao2025thematic}) separately.
The tool produced codebooks incrementally using two coder agents, one aggregator agent, and one reviewer agent.
The coders assign codes, the aggregator aggregates similar codes, and the reviewer performs the final review before producing the incremental codebooks.
The tool produces a codebook containing codes and quotes that support the development of our findings.

\section{RQ1 Results: Signal Collapse Mechanisms}
\label{sec:results-rq1}
Of the 252 articles in the corpus, 167 (66.3\%) documented a specific manipulation mechanism. 
Table~\ref{tab:mt-rq1} shows that the mechanism landscape is heterogeneous. Stars are overwhelmingly gamed by coordinated bot networks (MT-BOT in 31 of 32 Google Search sources).
Activity signals are gamed with both deliberate spoofing and legitimate AI inflation. 
Documentation is mostly degraded by AI-generated content, while community endorsement is gamed for both AI retrieval and human persuasion. 
Three signals: activity-legitimate (S3a), badges (S4), and issue-response (S5) are documented almost entirely outside software supply chain-specific literature (9\%, 0\%, and 6\% of the Google Search corpus are SSC-related, respectively), appearing instead in developer tooling and AI adoption research. 
Yet, those signals have previously been found in the literature to be used by developers when choosing dependencies~\cite{dabbish2012social,qiu2019signals,pashchenko2020qualitative,wermke2022committed,hamer2025trusting}.
This heterogeneity of mechanisms shows that no single intervention addresses the collapse across signals. 
The three findings that follow identify the patterns that cut across this diversity.

\begin{table}[]
    \centering
    \footnotesize
     \caption{Mechanism Technique Distribution across Trust Signals in the Google Search Corpus}
    \begin{tabular}{p{.43\columnwidth}|r|p{.4\columnwidth}}
        \hline
        Signal & n & Mechanisms \\ \hline
        \signal{S1a:stars} & 32 & 31 (97\%) MT-BOT, \\
        & & 1 (3\%) MT-BADGE-GAME \\
        \rowcolor{gray!20}
        \signal{S1b:downloads} & 25 & 9 (36\%) MT-BOT, \\
        \rowcolor{gray!20}
        & & 8 (32\%) MT-CRED,\\
        \rowcolor{gray!20}
        & & 6 (24\%) MT-PKG-INJ,\\
        \rowcolor{gray!20}
        & & 1 (4\%) MT-HALLUC, \\
        \rowcolor{gray!20}
        & & 1 (4\%) MT-AI-AGENT \\
        \signal{S2:documentation} & 8 & 6 (75\%) MT-DOC-FAKE,\\
        & & 2 (25\%) MT-HALLUC \\
        \rowcolor{gray!20}
        \signal{S3a:activity-legitimate} & 19 & 19 (100\%) MT-AI-AGENT \\
        \signal{S3b:activity-mimicry} & 46 & 21 (46\%) MT-SPOOF,\\
        & & 9 (20\%) MT-SOC-ENG, \\
        & & 6 (13\%) MT-DOC-FAKE, \\
        & & 4 (9\%) MT-CRED, \\
        & & 4 (9\%) MT-BOT, \\
        & & 1 (2\%) MT-PKG-INJ, \\
        & & 1 (2\%) MT-OTHER \\
        \rowcolor{gray!20}
        \signal{S4:platform-awarded-achievements} & 6 & 6 (100\%) MT-BADGE-GAME \\
        \signal{S5:issue-response} & 6 & 3 (50\%) MT-AI-AGENT, \\
        & & 2 (33\%) MT-PKG-INJ, \\
        & & 1 (17\%) MT-BOT \\
        \rowcolor{gray!20}
        \signal{S6:community-endorsement} & 25 & 13 (52\%) MT-UGC-PLANT, \\
        \rowcolor{gray!20}
        & & 4 (16\%) MT-SOC-ENG, \\
        \rowcolor{gray!20}
        & & 4 (16\%) MT-CRED, \\
        \rowcolor{gray!20}
        & & 4 (16\%) MT-HALLUC \\ \hline
    \end{tabular}
    \label{tab:mt-rq1}
\end{table}

% \subsection*{\newfinding{finding:inverted}{The economics of trust signals in the software supply chain have inverted: faking those signals now costs less than earning them.}}

\subsection*{\newfinding{finding:inverted}{Cheap trust signals in the software supply chain are collapsing because faking costs less than earning them, legitimate and adversarial actors produce identical signal outputs, and for some signals, the collapse is happening with no attacker involved at all.}}

Trust signals in the software supply chain were originally effective because producing them required genuine effort.
Building a project that attracted real users/downloads (\signal{S1}), maintaining it over time (\signal{S3}), and earning community recognition (\signal{S6}) all impose real costs on the trustee.
This cost asymmetry is what made those signals informative: legitimate projects could produce the signals more cheaply than impostors, creating a separating equilibrium between trustworthy and untrustworthy packages~\cite{spence1973}. 
Our corpora document that this asymmetry has collapsed. 

First, the cost of adversarial manipulation has fallen to commodity levels.
\signal{S1a:stars} are cheap~\doc{D1.1, D1.3, D1.4, D1.8, D1.10, D1.11, D1.17, D1.32, R12, R41, R51, R95, R102, R185, R191, R204, R494, R713, R768, R816}: they cost \$0.03--\$0.85 from publicly listed vendors; one Google Search source documents 63,872 suspected fake accounts through a single campaign.
Practitioners describe the collapse directly: \qt{GitHub stars used to really mean something... now they've been gamed by the dead internet just like everything else}~(Hacker News~\doc{D1.31}), with one independently naming the Goodhart dynamic: \qt{at some point stars were actually a great signal... but then once someone started using the signal to give dollars, the signal was compromised}~\doc{D1.31}.
\signal{S1b:downloads} are inflated through credential theft and automated install loops~\doc{D2.2, D2.3, D2.14, D2.18, D2.24, R494, R515, R549, R741, R806}. \signal{S2:documentation} that once required sustained effort is now generated by AI in seconds~\doc{D3.3, D3.4, D3.8, D3.12, D3.14, D3.17, D3.22}.
\signal{S3:activity} is also inflated: a practitioner in \texttt{r/cybersecurity} documented a 2-month campaign using 270+ commits to simulate legitimate development history before deploying malicious code~\doc{R477}. This pattern is the same trust-building pattern previously associated with high-value infrastructure attacks, now economically viable against smaller targets.

Second, 38.3\% of documented mechanisms in the Google Search corpus exploit techniques that are structurally indistinguishable from legitimate behavior at the signal level (MA-AMB). 
Soliciting stars from genuine community members (\signal{S1a}), optimizing content for AI retrieval (\signal{S6}), and using AI to generate documentation (\signal{S2}) are actions honest project owners also perform.
Thus, separating the adversarial and honest signals is difficult because the signals leave no trace of their motivation.

Third, 14.4\% of Google Search sources document non-adversarial structural inflation (MA-NON) with no deceptive actor involved. 
AI coding agents generate 275 million commits per week on GitHub, with commits rising 180\% while releases rose only 30\% \doc{D5.5, D5.6, D5.9, D5.10, D5.13, D5.31, R52, R206, R346, R373, R455, R465, R551, R638, R671, R841, R842}, inflating activity signals (\signal{S3a}) as a structural side-effect of legitimate development tooling.
These sources show that signals for contributor activity (\signal{S3a}), badges (\signal{S4}), and issue response (\signal{S5}) are collapsing even in the absence of any attacker.
Together, these three directions mean that even a perfect detector that can identify every bot account, coordinated campaign, and stolen credential would still face 52.7\% of the documented manipulation landscape in the Google Search corpus for which detection is not a meaningful concept.
The signals are not failing because attackers are too sophisticated to be detected, but because the same observable properties of those signals can be produced by honest effort, legitimate tooling, and deliberate manipulation alike, leaving no signal-level trace of their motivation~\cite{akerlof1970market}.\looseness=-1

\subsection*{\newfinding{finding:composite}{
The most sophisticated attacks do not manipulate a single signal; they reconstruct the entire appearance of a trustworthy project, rendering composite signal checks ineffective.}
}

When individual signal manipulation risks detection, the documented attacker response is not to abandon the manipulation but to manipulate more signals. 
The secondary mechanism distribution provides empirical evidence for this escalation: MT-PKG-INJ appears as a secondary code in 14.4\% of Google Search sources and consistently co-occurs with primary mechanisms, including MT-SOC-ENG and MT-DOC-FAKE, rather than appearing as a standalone technique.
The following campaigns are examples with multiple gamed signals. 

GitVenom~\cite{kaspersky2025gitvenom} combined fabricated README files, spoofed commit histories, and package injection to simultaneously manipulate \signal{S2:documentation} and \signal{S3b:activity-mimicry}~\doc{D3.1, D3.2, D3.3, D3.4, D3.10, D3.11}.
The ESLint compromise~\cite{eslint2018compromise} manipulated three signals via a single campaign: credential theft (\signal{S3b}), phishing-harvested maintainer identity (\signal{S3, S6}), and malicious package publication (\signal{S1b})~\doc{D6.16, D6.17, D6.20, D6.22}.
The xz utils attack~\cite{freund2024xz} invested two years of genuine commit activity (\signal{S3b}) and sustained community participation (\signal{S6}) before deploying its payload~\doc{D6.1, D6.7, D6.8, R98, R100, R108, R110, R112, R157, R196, R220, R237, R331, R549, R593, R673}.
The May 2026 Fedora incident extended this pattern to AI-assisted execution: an agent under compromised credentials simultaneously manipulated \signal{S3:activity}, \signal{S5:issue-response}, and \signal{S2:documentation} with justifications that ``eventually overwhelmed the maintainer into merging the fix''~\cite{lwn2026fedora}. 
A practitioner in \texttt{r/netsec} observes the same envelope-reconstruction logic applied to mobile libraries: \qt{Malicious SDKs [backdate] their commit history to look established, adding fake stars and forks, then getting recommended in top Android libraries' blog posts}~\doc{R541}, gaming \signal{S3:activity}, \signal{S1a:stars}, and \signal{S6:community-endorsement} simultaneously.
In each case mentioned above, the attacker did not game a single signal; they operated across every dimension that a composite evaluator would check.

\section{RQ2 Results: Responses to Signal Collapses}
\label{sec:results-rq2}
Of the 252 sources in the Google Search corpus, 163 (64.7\%) contain a practitioner, platform, or researcher voice on how the ecosystem has responded to trust signal manipulation. 
Table~\ref{tab:rq2} shows the distribution of response types across three dimensions in the Google Search corpus: (1) statement type, which captures whether the response is advice only or describes an action actually carried out; (2) trust direction, which captures what the source records of the practitioner's reliance on the signal; and (3) proposed signal alternative. 
The four findings that follow build on this distribution to identify the structural gaps in the ecosystem's documented response.

\begin{table}[]
    \centering
    \footnotesize
    \caption{RQ2 Responses Distribution in the 252 Google Search Corpus}
    \begin{tabular}{p{0.75\columnwidth}|rr}
        \hline
        Code & \# & \% \\
        \hline
        \multicolumn{3}{l}{\textbf{Statement type}}\\ \hline
        A — Stated only & 89 & 54.6\% \\
        D — Analyst-inferred & 44 & 27.0\% \\
        C — Coupled & 26 & 16.0\% \\
        B — Revealed only & 4 & 2.5\% \\ \hline
        \multicolumn{3}{l}{\textbf{Trust direction}} \\ \hline
        Signal downweighted & 90 & 55.2\% \\ 
        Signal replaced & 38 & 23.3\% \\
        Signal reaffirmed & 21 & 12.9\% \\
        No change & 3 & 1.8\% \\
        Signal abandoned & 2 & 1.2\% \\
        \hline
        \multicolumn{3}{l}{\textbf{Proposed alternative signal/s (n=49)}} \\ \hline
        Cheap observable signals (fork ratios, issue quality, commit history, contributor activity) & 32 & 65.3\% \\
        Expensive / hard-to-fake signals (commit signing, provenance, SBOM, attestation) & 17 & 34.7\% \\
        \hline
    \end{tabular}
    \label{tab:rq2}
\end{table}

\subsection*{
\newfinding{finding:advice}{
The documented response to trust signal collapse is predominantly advice without action: 54.6\% of Google Search sources contain advice about what practitioners should do, while only 1 in 6 documents an action that was actually carried out.
}}

Of the 163 Google Search sources with a documented response, 54.6\% contain advice or stated concerns without any actual enacted action (statement type A).
These documents contain researchers/article writers recommending signal substitution/practices, vendors advising composite evaluation, or practitioners expressing concern in forum threads. 
In the Google Search corpus, the advice clusters into four categories: composite evaluation (combining multiple signals), tooling adoption (automated scanners, dependency checkers, and SCA tools), process controls (manual code review before adoption and delayed updates), and platform escalation (abuse reporting and waiting for platform enforcement).
All those advice tend to suggest adding verification around the signal without displacing the signal itself.
An example from a practitioner on Reddit \texttt{r/github}: \qt{Just assume that anything on GitHub is malware unless you know and trust the repository owner, or you've done at least some basic sanity-checking of the code. Stars have never meant anything.}~\doc{R191}.
A further 27.0\% of Google Search sources are analyst-inferred (type D): the coder reads a response from indirect evidence in the source rather than from an explicit practitioner statement, e.g., websites to get downloads count (\signal{S1b}). 
Only 18.4\% of Google Search sources document an action that was actually carried out (B (action only) + C (stated + action)). 
This pattern holds across signals: \signal{S2:documentation} records 0\% B or C responses in the Google Search corpus despite documenting the high experimental attack success rates in the corpus (84\% via README injection \doc{D3.17}). \signal{S5:issue-response} also records no described actions at all in the Google Search corpus. 
The literature on responding to cheap signal collapse is more a literature of advice than of documented behavior change.

% \subsection*{
% \newfinding{finding:downweigthed}{
% % Documented responses mostly reduce reliance on a trust signal rather than replace or abandon it, suggesting that gameable signals remain embedded in dependency evaluation despite acknowledged unreliability.
% }}

\subsection*{
\newfinding{finding:downweighted}{
Documented responses show that practitioners reduce, rather than replace, reliance on trust signals, but the added verification overhead increases the cost of those signals.
}}

When a response is documented, the dominant pattern is not exit or substitution but downweighting: continuing to consult the signal while trusting it less. 
Fifty-five percent of Google Search sources with a documented trust direction record signal downweighting, against 23.3\% that describe signal replacement and 1.2\% that describe abandonment. 
% This pattern is consistent across statement types: among A-only Google Search sources (prescriptions without described action) 54\% show signal downweighting as the recorded trust direction, meaning the advice itself is oriented toward using the signal more carefully rather than discarding it. 
Rather than removing unreliable signals from the evaluation workflow, sources across both corpora advise adding verification layers around them: automated scanning~\doc{D2.2, D2.11, R2, R51, R62, R92, R104, R159, R177, R371, R532, R538, R541, R549, R554, R579, R626, R632, R659, R715, R716, R717, R721, R727, R738, R743, R752, R754, R757, R835, R869}, composite evaluation across multiple signals~\doc{D1.30, D2.3, D2.11, R613, R635, R713, R770, R771, R783, R826, R854}, delayed updates to observe community response~\doc{R593, R673, R757}, and manual inspection of contributor histories~\doc{D1.2, D1.13, R98, R156, R532, R613, R660, R696, R719, R757, R759, R870}.\looseness=-1
The cost of this overhead is documented directly: practitioners report spending more time evaluating dependencies than writing code, and institutional responses include maintaining internally audited mirrors at high organizational cost~\doc{D2.27, D9.1, D9.2, R824, R843, R858}.
This verification overhead is caused by the same AI-driven volume that degrades the signals: \qt{If AI increases code volume by 10x, human review becomes a fatal bottleneck} (\texttt{r/LocalLLaMA}~\doc{D5.12}). 
When the bottleneck is breached, \qt{security review is becoming an afterthought in AI-driven development} (\texttt{r/ExperiencedDevs}~\doc{D5.30}).
Thus, \textbf{the signal remained cheap to fake but became expensive to trust}, inverting the original transaction cost advantage of open-source reuse~\cite{arrow1974limits}: what began as a zero-verification adoption decision now requires verification infrastructure that rivals the cost of the dependency itself.
% "A pipe can only hold so much volume. Right now, you have pressure on the input side, and it's more like a stormwater drain than a filtered water outlet" (r/LocalLLaMA). 

A minority of sources document the logical alternative to this cost escalation: bypassing signals entirely in favor of direct inspection, i.e., reading source code before adoption \doc{D1.2, D3.4, D3.5, D3.11, D3.22, D6.3, R141, R383, R757}, auditing maintainer contribution histories across repositories \doc{D6.44, R141, R757, R824}, or running packages in a sandbox before integration \doc{D1.30, R371, R383, R532, R706}.
A practitioner mentioned in \texttt{r/opensource}: \qt{Normally, the idea isn't that you trust open-source: you have all the code, you can look at it.}~\doc{R383}.
Appearing in only a minority of both corpora suggests that this approach remains a minority practice rather than a widely documented or institutionalized response, possibly due to its high cost.
% These responses sit at the opposite end of the cost curve from downweighting: they are individually rational but scale poorly, and they are documented almost exclusively by experienced practitioners rather than researchers or vendors, suggesting they exist as tacit practitioner knowledge rather than institutionalised practice.
Among Google Search sources where a practitioner describes an actual action (B or C), the same conservatism holds: signal replacement and downweighting split nearly evenly in type C sources, and abandonment appears in only 1 of 30 described-action sources.
Signal reaffirmation is documented in 12.9\% of Google Search sources that mention the persistent use of signals, though most are analyst-inferred rather than author-stated, suggesting that continued use is more often observed than actively defended.
The corrupted signals are not discarded; they are retained, adding verification overhead and leaving the trust evaluation infrastructure largely intact while reliability declines and costs rise.

\subsection*{
\newfinding{finding:responses}{
The proposed responses to signal collapse: substituting one signal for another and aggregating several together, are largely self-defeating: the proposed substitutes are themselves documented cheap gameable signals, and aggregating signals offers only short-term protection against a patient attacker.
}}

Of the 163 Google Search sources with a documented response, 49 (30.1\%) propose a specific alternative signal, 65.3\% of which are cheap, observable metrics: fork-to-star ratios, issue quality, and commit frequency. 
% Those signals fall into the same manipulability category as the ones they replace.
% Of the 163 Google Search sources with a documented response, 51 (30.5\%) propose a specific alternative signal. 
% Of these, 64.2\% propose cheap observable metrics: fork-to-star ratios, issue quality, contributor lists, commit frequency, last commit date.
% These signals are observable without cryptographic infrastructure and sit in the same manipulability category as those they are proposed to replace. 
This substitution response is directly visible in the Google Search corpus: \signal{S3:activity} is the most commonly proposed substitute for \signal{S1a:stars}~\doc{D1.2, D1.4, D1.10, D1.13, D1.22, D1.27, D1.28, D1.30, R137, R141, R191, R372, R537, R748, R757, R824, R866}, e.g., \qt{look at commits/contributors instead of stars}.
% yet \signal{S3a:documents} documents 19 sources recording AI coding agents inflating contributor activity at scale with no adversarial intent, and \signal{S3b} documents 46 sources recording deliberate spoofing of the same signal.
The proposed substitutes are already failing by the same mechanisms as the signal they replace.
% These signals are observable without cryptographic infrastructure and sit in the same manipulability category as those they are proposed to replace. 
% The circular logic is visible in the corpus directly, e.g., \signal{contributor activity} is the most commonly proposed substitute for \signal{star counts}.
% Yet \signal{S3a:activity-legitimate} documents 19 sources recording AI coding agents inflating contributor activity at scale with no adversarial intent, and \signal{S3b:activity-mimicry} documents 46 sources recording deliberate spoofing of the same signal. 
The aggregation response~\doc{D1.22, D1.24, D1.28, D1.30, D2.3, D2.35, D3.31, R613, R715, R770, R771, R783, R854}, e.g., checking \signal{S1a:stars}, \signal{S2:documentation}, \signal{S3:activity}, and \signal{S6:community endorsement} together, rests on the assumption that manufacturing multiple signals simultaneously is prohibitively costly.
This assumption has been refuted in Finding~\ref{finding:composite}: the same corpora that advise composite evaluation also document coordinated campaigns that reproduce the full signal envelope simultaneously, which defeat the composite check by construction.
% Finding~\ref{finding:composite} documents that this assumption does not hold for a patient attacker: the same Google Search corpus that prescribes composite evaluation also documents coordinated campaigns that reproduce the full signal envelope simultaneously, defeating the composite check by construction.
Aggregation raises the cost of manipulation in the short term but does not change the structural properties of the signals being aggregated.

% The aggregation response fares no better: there are sources prescribing composite evaluation, e.g., checking \signal{stars (S1)}, \signal{documentation (S2)}, \signal{activity (S3)}, and \signal{community engagement (S6)} together.
% This aggregation does not acknowledge that a patient attacker reproducing all signals simultaneously, as documented in Finding~\cite{finding:composite}, defeats the composite check by construction. 

Only 34.7\% of Google Search sources propose alternative signals with cryptographic infrastructure: \signal{commit signing}, \signal{SBOM}, \signal{SLSA attestation}, \signal{Sigstore provenance}, that would structurally resist manipulation even with a higher cost. 
These signals are also proposed by some practitioners in Reddit~\doc{R81, R93, R98, R286, R425, R455, R477, R488, R505, R510, R514, R543, R605, R660, R784, R803}. A practitioner mentioned using \qt{verifiable trust signals — not stars or hype. The signals include OSSF Scorecard, build provenance (SLSA), signed commits, license transparency, and maintenance patterns.}~\doc{R803}.
Commit signing is the single most frequently proposed alternative (8 Google Search sources), yet is adopted by only 10\% of GitHub repositories~\cite{sharma2025prevalence}, leaving the most structurally sound proposed remedy as rare in practice as it is sound in principle.
The gap between proposal and adoption suggests that cryptographic infrastructure, while theoretically sound, faces barriers to uptake.

% One cheap signal is used as a substitute for another, e.g., contributor activity to substitute for stars, yet both can be gamed.

% Some recommend aggregating different cheap signals, e.g., using both activity and popularity.

% \subsection*{
% \newfinding{finding:non-adv}{
% Non-adversarial signal inflation, i.e., legitimate AI tooling degrading the same signals as adversarial attacks, lacks documented responses with actual behavior change, suggesting the ecosystem's response infrastructure remains calibrated for malicious actors and has not yet adapted to this legitimate vector.
% }
% }

\subsection*{
\newfinding{finding:non-adv}{
Non-adversarial signal inflation (i.e., by legitimate AI tooling) has elicited only a limited number of documented responses that result in actual behavior change, and none at the institutional level, suggesting that the ecosystem's response infrastructure has not yet adapted to this legitimate vector.
}}

The Google Search corpus contains 24 sources in which signal degradation is classified as non-adversarial (MA-NON), e.g., \signal{activity} inflation caused by legitimate AI coding agents and CI/CD infrastructure, and \signal{downloads} inflation caused by mirror systems operating without deceptive intent. 
In one of these sources, a package maintainer, independently investigating download statistics, found that \qt{most people downloading my package were things like `z3c.pypimirror/1.0.15.1' and `pep381client' } (StackOverflow~\doc{D2.24}). 
This mirror infrastructure inflates counts before any adversarial actor acts.
Of the 24 Google Search sources classified as MA-NON, 33\% have no documented response of any kind. 
The 67\% that do contain a response are predominantly general advice (type A, n=9) or analyst inferences (type D, n=4), with only 3 documenting actual behavior change.
% show a distinct pattern from adversarially manipulated signals: the majority are general prescriptions (statement type A, n=11) or analyst inferences (type D, n=9), with only 6 Google Search sources documenting actual described behavior change (B or C).
% Where responses are described, they are individual practitioner adaptations — monitoring actual package usage against phantom downloads~\doc{S2.14}, replacing download signals with community activity metrics~\doc{S1.21}, abandoning repositories with artificially inflated stars~\doc{S1.14} — rather than institutional or platform-level countermeasures.
The contrast with the adversarial subset is direct: MT-BOT Google Search sources (n=55) elicit concrete advice responses: abuse-reporting pipelines, automated detection tools, and platform governance escalation
% MT-AI-AGENT sources (n=23) attract observations about volume growth and concerns about signal reliability, but no documented countermeasure at the practitioner level. 
The detection infrastructure documented in both corpora, e.g., StarScout's lockstep signature analysis \doc{D1.1, D1.11, R715}, ghost-account ratio heuristics \doc{D1.12, D1.17, R715}, and behavioral fingerprinting \doc{D8.6, D8.13, R770}, requires an adversarial actor to detect. 
Non-adversarial inflation produces activity signatures indistinguishable from genuine high-productivity development, leaving no signal-level trace that existing tooling is designed to flag. 
A measurement constraint theorized directly in the corpus: \qt{when we detect agentic activity, we can never be sure of the proportion of the activity that is due to the agent, due to the Peril of Partial Observability}~\doc{D5.15}.
The response gap is therefore not a matter of practitioner awareness but of infrastructure: the ecosystem's verification tooling was designed for a threat model that excludes legitimate tools as a degradation vector.
% The response gap is therefore not a matter of practitioner awareness (the AI-driven inflation is well-documented), but of infrastructure: the ecosystem's verification tooling was designed for a threat model that does not include legitimate tools as a degradation vector.
% This detection difficulty is theorized directly in the Google Search corpus: researchers note that \qt{when we detect agentic activity, we can never be sure of the proportion of the activity that is due to the agent, due to the Peril of Partial Observability}~\doc{D5.15}. 
% This limitation shows a measurement constraint that existing detection tools are not designed to address.

% The AI-driven inflation of \signal{commit counts} and \signal{activity} signals is well-documented in the corpus — 
\section{Limitations}
\label{sec:limit}

We report limitations using established criteria for qualitative trustworthiness~\cite{lincoln1985} to help readers calibrate their inferences.

\noindent \textbf{Corpus completeness and subjectivity.}
The Google Search corpus is bounded by the retrieval strategy: sources not indexed by Google Search or the practitioner platforms we queried are automatically excluded.
Non-English-language practitioner communities and private organizational grey literature are underrepresented.
The corpus reflects a retrieval window ending on July 19th, 2026. We acknowledge that the manipulation landscape and ecosystem responses of signals in the software supply chain are evolving rapidly; thus, findings specific to tool names, pricing figures, and adoption rates may become outdated.
The 870 Reddit threads were retrieved using an automated tool that applied LLM-based relevance filtering; threads judged borderline relevant were excluded regardless of content.
The corpus covers threads from 2023 to 2026 and is bounded by subreddit coverage: communities not indexed by the retrieval tool, non-English subreddits, and deleted or private threads are absent.
On subjectivity, the corpus consists predominantly of experiential and self-reported accounts rather than structured evaluations, filtered through each source's own framing and incentives.
The coding captures what sources \textit{record} as mechanisms and responses, not independently verified ground truth.

\noindent \textbf{Signal coverage.}
In the Google Search corpus, the \signal{S2:documentation} (n=14) and \signal{S4:platform-awarded-achievements} (n=11) are thin relative to other signals.
Signal-specific findings for these two categories are less generalizable than those for \signal{S1a}, \signal{S1b}, and \signal{S3b}, which are substantially better evidenced.
% The small size of \signal{S2} and \signal{S4} is acknowledged as a gap the corpus surfaces rather than resolves.

\noindent \textbf{Coding reliability.}
Intercoder reliability was assessed by comparing first-author manual codings against LLM codings on a 40-source sample after three rounds of prompt calibration.
All RQ1 closed-code fields exceeded 70\% agreement.
We acknowledge the subjectivity of the single coding judgment for the \textit{trust\_direction} field (RQ2); yet the narrow decision space limits this subjectivity.
The coding for the Reddit corpus was conducted by multiple LLM agents without manual validation of individual codes. The produced codebook for this Reddit corpus reflects machine-identified codes rather than first-author-verified codes. Yet, the included Reddit IDs/quotes are manually verified to ensure they are relevant before being incorporated into the paper.

\noindent \textbf{Construct validity and Generalisability.}
The RQ2 coding captures what the literature \textit{records} as ecosystem responses, rather than directly observed actions.
Sources describing responses are filtered through their authors' framing choices and may not reflect the full range of practitioners' behavior.
% The prescription-action gap documented in Finding~\ref{finding:advice} reflects the documented literature; the actual gap between advice and adoption in practitioner behaviour requires primary empirical investigation to establish.
% \noindent \textbf{Generalisability.}
The Google Search corpus comprises eight signals selected based on prior empirical work on dependency adoption decisions.
Other observable signals that practitioners could consult, e.g., license type, organizational affiliation, vulnerability scan results, are outside the scope.
The findings characterize the range of documented mechanisms and responses observed across both corpora rather than the whole population distribution.
Quantitative findings (percentages, counts) describe the Google Search corpus and should not be read as ecosystem-wide prevalence.

\section{Discussions, Implications, and Conclusions}
\label{sec:discussion}
The findings paint a coherent picture of cheap trust signal collapses across eight signals used in the software supply chain.
The documented response is predominantly advice without enacted action, and the proposed substitutes recycle other cheap signals that are also failing. 
This collapse describes an ecosystem drifting toward the condition Akerlof predicted for any market in which quality signals lose their informational content: a market for lemons~\cite{akerlof1970market}.
In a market for lemons, the inability to distinguish good packages from bad erodes the incentive to produce good ones and the willingness to trust any of them.
In this section, we discuss why the present trajectory is unsustainable, why the two most commonly documented response strategies (signal substitution and manipulation detection) cannot fully break the cycle, and why the structurally sound alternatives identified in the literature are known but not yet adopted at scale.

\subsection{Why the Current Trajectory Is Unsustainable}
In Section~\ref{sec:introduction}, we established that cheap trust signals only work when the cost of faking them is more expensive than the cost of earning them honestly, and that the collapse of these signals produces a market for lemons in which good trustworthy dependencies become indistinguishable from the untrustworthy ones.
The findings provide empirical evidence that this collapse is not a future risk but \emph{a present trajectory}, visible in the response patterns documented in the corpus.\looseness=-1

% The prescription gap (Finding~\ref{finding:advice} (54.6\% of responses are stated only) suggests the lack of actual adoption of costlier signals or verification that involves cost.
Finding~\ref{finding:downweighted} shows that 55.2\% of documented trust direction in the Google Search corpus is downweighting signals, suggesting that informed actors are quietly reducing reliance without completely exiting. This response is a rational response when the signals are gameable, yet verifying everything is not possible.
Another part of Finding~\ref{finding:downweighted} shows that 12.9\% of documented trust directions in the Google Search corpus are reaffirming signals, suggesting that uninformed/unaware actors have not yet updated their prior strategies for choosing dependencies and still rely on gameable signals.
These two patterns are not independent: they are the same lemon dynamic at different stages of awareness. 
Together, those responses describe an ecosystem in which the most informed participants are quietly reducing reliance on cheap signals while the least informed continue to rely on them.
Many stated-only Google Search sources advise platform abuse reporting, community flagging mechanisms, and automated detection tools intended to increase practitioners' awareness.
Whether these institutional advice translate into adopted practice is precisely the gap Finding~\ref{finding:advice} documents: the advice exists, but the behavior change does not yet follow at scale.

Finding~\ref{finding:inverted} documents that mimicry has become cheap enough to be industrialized, openly marketed, and calibrated against published detection signatures.
This mimicry is known in biology as Batesian mimicry: when cheap mimics free-ride on an honest signal until it inflates and loses its discriminating power entirely~\cite{searcy2005evolution, spence1973}.
% When mimicry is this cheap, the single-crossing condition that makes a signal informative is violated~\cite{spence1973}: low-quality (or even malicious) dependency maintainers can produce the signal at a cost that no longer distinguishes their dependencies from those of high-quality maintainers.
% The separating equilibrium collapses into a pooling equilibrium, and the signal ceases to carry information about quality regardless of how carefully it is read.
% The separating equilibrium collapses, and the cost rises.
% Arrow's characterization of trust as the lubricant of economic systems~\cite{arrow1974limits} suggests that high-trust regimes reorganize under pressure rather than dissolve.
The consequence is rising costs: reorganization under signal collapse means privatizing the verification cost that the signal ecosystem once distributed~\cite{arrow1974limits}.
% But reorganization under these conditions means privatizing the verification cost that the signal ecosystem once distributed.
% Practitioners spend more time evaluating dependencies; organizations maintain internal mirrors and curated dependency lists that replicate at private expense what the ecosystem once provided as a public good; and security teams absorb overhead that scales with the number of dependencies rather than with their actual risk.
Finding~\ref{finding:downweighted} documents this rising cost directly: practitioners report spending more time evaluating dependencies than writing code, and institutional responses include maintaining internally audited mirrors at significant organizational cost~\doc{D2.27, D9.1, D9.2, R824, R843, R858}.
% Finding~\ref{finding:downweighted} documents this rising cost directly: sources describe spending more time evaluating dependencies than writing code, and institutional responses include maintaining internally audited mirrors at significant organizational cost~\doc{D2.27, D9.1, D9.2, R824, R843, R858}.
% As Finding~\ref{finding:substitution} shows, proposed substitutes do not reduce this overhead — they displace it onto a different set of signals that will attract the same manipulation pressure once they become consequential evaluation criteria.
The Reddit corpus adds a further lock-in dimension: switching costs trap practitioners in existing dependencies regardless of signal reliability: \qt{a simple calculus of what you gain versus the effort required to move — actually almost none}~\doc{R64} (\texttt{r/Python}).
This cost difference suggests that the ecosystem could not self-correct through individual exits even when practitioners are aware of the collapse.
The net value of open-source reuse erodes as verification overhead accumulates without a structural resolution.\looseness=-1
% The net value of open-source reuse, i.e., its transaction cost advantage over proprietary alternatives, erodes as verification overhead accumulates without a structural resolution.
% The Reddit corpus documents a further dimension of this trap: switching costs lock practitioners into existing dependencies regardless of signal reliability. 
% As one practitioner notes, the decision not to migrate is \qt{a simple calculus of what you gain versus the effort required to move — actually almost none}~\doc{R64} (\texttt{r/Python}). The consequence is that the ecosystem cannot self-correct through individual practitioner exits even when those practitioners are aware of the collapse.
% The cost of exiting a widely used dependency exceeds the cost of continuing to accept the risk of unreliable signals.

\noindent\textbf{Implication for Platform Operators.} 
PyPI's removal of download count~\cite{pypidownloads} displays is the only documented instance of a platform retiring a gameable signal rather than continuing to surface it (Finding~\ref{finding:advice}). 
Extending this platform-level signal retirement to other signals and registries could be the structural action the corpus most directly supports.

\subsection{Why Signal Substitution and Aggregation Cannot Break the Cycle}

Finding~\ref{finding:responses} documents that the literature's two proposed responses are to substitute a failing signal with a different one, or aggregate several signals together.
Both responses have the same structural vulnerability.
Substitution fails because the proposed alternatives are also predominantly cheap, gameable signals, and aggregation fails because Finding~\ref{finding:composite} shows that patient attackers can reproduce the multiple signals simultaneously.
To understand why both substitution and aggregation are structurally insufficient, we examine the conditions identified in the signaling literature under which signals can survive manipulation pressure~\cite{searcy2005evolution,spence1973}.

\textit{Shared interest} keeps signals honest when the signaler's payoff is coupled to the receiver's welfare.
In software, this condition holds within funded maintainer teams and organizations with liability exposure, but evaporates the moment the dependency graph becomes anonymous and transitive.
XZ utils~\cite{freund2024xz} documents the failure mode: years of simulated alignment harvesting, in which personal trust communities extend to apparently engaged contributors~\doc{D6.7}.
The cost of producing that simulated alignment is further reduced by the emergence of AI.
\textit{Genuinely costly signals} impose a cost disproportionately on low-quality producers.
This condition has been violated for cheap signals by various mechanisms documented in RQ1 (\S\ref{sec:results-rq1}), including AI.
Any signal whose cost of producing plausible-looking information is now nearly zero~\cite{spence1973,goodhart1975problems}: a thoughtful README, a high-volume commit history, a contextually appropriate issue response.
The substitution response in Finding~\ref{finding:responses} is this dynamic applied recursively: \signal{S3:activity} is proposed as a substitute for \signal{S1a:stars}, while both corpora document that \signal{S3:activity} is collapsing for identical reasons.
Under Goodhart's Law~\cite{goodhart1975problems}, any signal that becomes a target ceases to be a good measure: once practitioners rely on a signal for adopting dependencies, the signal becomes worth faking, its cost falls, and it stops separating trustworthy packages from untrustworthy ones.
The collapse, which rendered the original signal unreliable, also occurs for the substitute.
Aggregation does not escape this logic: combining signals that can each be individually manufactured offers only short-term protection, and Finding~\ref{finding:composite} documents that sophisticated attackers already manufacture all components simultaneously.

\textit{Detection and punishment} deter mimicry when the probability of being caught, multiplied by the cost of being caught, is high.
In open-source ecosystems, both terms collapse.
The probability of being caught is getting lower as 38.3\% of documented mechanisms in the Google Search corpus are indistinguishable from legitimate behavior (Finding~\ref{finding:inverted}). 
The cost of being caught is reduced as identity is disposable: an exposed actor can create a fresh pseudonym at zero cost (as in xz utils attack~\cite{freund2024xz}), so punishment does not accumulate.

What survives this analysis is a narrow but constructive residual: signals whose production is \textit{causally bound} to the quality they certify; equivalent to indices rather than signals in the biological literature~\cite{searcy2005evolution}.
A cryptographic signature cannot be produced without the private key; a reproducible build cannot be forged without controlling the source.
These signals are costly and cannot be eroded by AI; they require producers to do the real thing.
Finding~\ref{finding:responses} documents that these are the only proposed alternatives that would structurally resist the substitution collapse, appearing in 34.7\% of proposed alternatives but adopted in only 10\% of repositories, even for the most commonly proposed option.
% The corpus identifies the exit from the treadmill but cannot document the ecosystem taking it.

\noindent \textbf{Implications for Practitioners.}
Adopting signals whose production is causally bound to the quality they certify: commit signing, SBOM, and provenance attestation impose costs that cannot be eroded by AI.
Those signals require cryptographic infrastructure rather than the production of plausible-looking information. 
Where these signals are not yet available, sustained personal relationships with maintainers (e.g., known contributors and organizations with public accountability) represent the \textit{shared-interest mechanism} that both theory and the corpus identify as more manipulation-resistant.

\subsection{Why Detection Cannot Break the Cycle Either}
% The previous subsection established that signal substitution and aggregation are structurally insufficient, as AI has collapsed the cost of producing plausible-looking information.
% A natural response is to ask whether better detection can compensate: if we cannot replace cheap signals with more reliable ones, perhaps we can identify which cheap signals have been manipulated and discount them accordingly.
Signal substitution and aggregation cannot compensate because AI has collapsed the cost of producing plausible-looking information. 
A natural follow-up is whether better detection can help: if cheap signals cannot be replaced, perhaps manipulated ones can be identified and discounted.
% Together, these account for 52.7\% of the documented collapse landscape — a majority for which detection calibrated for adversarial intent is not a meaningful concept, because there is no adversarial intent to detect.
% \subsubsection*{The dual-use problem}
Finding~\ref{finding:inverted} documents why this response also faces structural limits.
First, 38.3\% of documented mechanisms in the Google Search corpus (MA-AMB) use techniques indistinguishable from legitimate behavior, e.g., soliciting stars from real users and AI-generated documentation are all actions honest project owners also perform, leaving no signal-level trace of intent~\cite{spence1973}.
Second, 14.4\% of mechanisms (MA-NON) involve no adversarial actor at all: AI coding agents inflate activity signals as a side-effect of legitimate development, and mirror systems inflate download counts without deceptive intent.
Detection infrastructure designed to identify manipulation lacks a mechanism to address the collapse caused by adopting legitimate tools.
Finding~\ref{finding:non-adv} documents that non-adversarial signal inflation lacks documented responses with actual behavior change, showing that the response infrastructure was built for a different threat model.
% — not because practitioners are unaware of it, but because the response infrastructure was built for a different threat model.
% The corpus documents that even measuring the scale of non-adversarial inflation is currently difficult: standard attribution methods recover only 3.3% of AI coding agent commits and miss 79% of AI agent activity~\doc{S5.2, S5.4}, suggesting that detection infrastructure designed for intentional manipulation is poorly positioned to address structural inflation.

\noindent \textbf{Implications for Tool Builders.}
Automated verification tools reduce manual verification cost and are widely documented as a concrete response (Finding~\ref{finding:advice}). 
However, existing tools are predominantly calibrated for adversarial manipulation.
The corpora document that even measuring the scale of non-adversarial AI-driven inflation is currently difficult: standard attribution methods recover only 3.3\% of AI coding agent commits and miss 79\% of AI agent activity~\doc{D5.2, D5.4, R638}.
This struggle suggests that detection infrastructure designed for intentional manipulation may be poorly positioned to address non-adversarial signal degradation~(Finding~\ref{finding:non-adv}).
A complementary class of tooling that measures signal \textit{reliability} rather than signal \textit{manipulation} may represent a productive direction that the corpus points toward. 
These tools should be able to flag statistical anomalies (e.g., in commit-to-release or download-to-dependent ratios) regardless of adversarial intent.

\noindent \textbf{Implications for Platform Operators.} 
The non-adversarial vector has no platform-level documented response in the corpus. 
The most intuitive structural remedy is transparent labeling of AI agent activity at the platform level, i.e., distinguishing AI-generated commits and contributions from human ones in the signal display.
Recent research demonstrates that AI coding agents exhibit detectable behavioral fingerprints, achieving an F1 score of 97.2\% across five major agents~\cite{ghaleb2026fingerprinting}, suggesting that platform-level attribution is technically feasible. 
Community conventions such as \texttt{Co-authored-by} tags and emerging kernel attribution guidelines~\cite{linuxkernel2026} represent early steps in this direction, but no package registry or code hosting platform has yet systematically disclosed AI contribution rates as part of its signal infrastructure. 
This transparency does not detect manipulation but removes the ambiguity that makes non-adversarial inflation invisible to downstream users.

\noindent \textbf{Implications for Researchers.} 
The non-adversarial vector (MA-NON) is documented in 40 Google Search sources but elicits only 6 described behavior changes, all of which are individual practitioner adaptations rather than institutional responses~(Finding~\ref{finding:non-adv}).
The specific impact of AI agent adoption on \signal{S4:platform-awarded-achievements} and \signal{S5:issue-response} as trust signals for dependency adoption decisions remains largely unexplored.
While the corpora document these signals inflating as a side effect of AI tooling, they do not document whether or how practitioners are adjusting their reliance on them in response to adopting these dependencies.
Empirical studies of how practitioners read and respond to AI-inflated signals represent the most direct future work this corpus surfaces.

\subsection{Why the Known Remedies are Not Being Adopted}
Finding~\ref{finding:responses} documents that the structurally sound alternatives, i.e., cryptographic attestation, provenance signing, and reproducible builds, appear in 34.7\% of proposed responses but are adopted by only 10\% of repositories, even for the most commonly cited option.
The previous discussion established what these alternatives are and why they would work.
Yet, we have not discussed why the adoption gap persists when practitioners already know the remedy.
The answer is not ignorance: the corpus documents awareness without adoption across multiple signals and multiple years (Finding~\ref{finding:advice}).

The barrier is structural: cryptographic attestation adopted by 10\% of repositories does not restore the separating equilibrium. This adoption instead creates a two-tier market in which sophisticated practitioners use costlier signals while the majority cannot distinguish attested from unattested packages, leaving the lemon dynamic intact~\cite{akerlof1970market}.
Attestation has network effects: its value is realized only when absence becomes a red flag~\cite{farrell1985standardization}, so early adopters bear the full cost while the ecosystem-level benefit remains unrealized.
This barrier explains why the corpora document the same alternatives proposed for years, yet adoption is scarce: recommendations do not change the incentive to adopt, as Olson predicts; coordinated change requires coercion or selective incentives~\cite{olson1965logic}.
The historical record confirms that excess cost is resolved by changing the default for all participants simultaneously~\cite{farrell1985standardization}.
The software supply chain equivalent is partially institutionalized (e.g., the EU Cyber Resilience Act and PyPI's trusted publishing) but remains limited in reach.
From an Arrow trust-as-lubricant perspective~\cite{arrow1974limits}, the goal is not to maximize individual verification effort but to restore the shared institutional infrastructure that makes cheap reuse safe again.\looseness=-1

\noindent\textbf{Implication for Policy Makers.} 
The theoretical analysis above explains why the adoption gap documented in Finding~\ref{finding:responses} has persisted despite repeated recommendations.
In adjacent domains, this problem was resolved by making the costlier signal the platform baseline rather than a voluntary option, e.g., Sender Policy Framework (SPF)/DomainKeys Identified Mail (DKIM) for email authentication~\cite{rfc7208}.
% The historical precedents in adjacent domains, e.g., Sender Policy Framework (SPF)/Domain Keys Identified Mail (DKIM) for email authentication~\cite{rfc7208}, succeeded not through voluntary adoption but by making the costlier signal the platform baseline: the index was enforced as a precondition for participation rather than an option practitioners could ignore.
The software supply chain equivalent is platform-level enforcement of signed provenance for packages above a usage threshold.
This enforcement makes the use of costlier signals no longer a recommendation but a precondition for registry publication or inclusion, which changes the default for all participants simultaneously rather than relying on individual opt-in.
Policy makers could extend the scope of mandate-level instruments beyond federal contractors by broadening coverage to widely adopted open-source dependencies.
% , which is the policy lever most directly supported by both the corpus evidence and the collective action analysis.
% The corpus documents signing and provenance attestation as proposed responses that shift verification costs upstream to producers rather than downstream to consumers (Finding~\ref{finding:substitution}). 
% The gap between their documented proposal rate (34.7\% of proposed alternatives) and their documented adoption rate suggests that mandate-level instruments could be used to achieve the ecosystem-wide adoption at which these mechanisms become effective.

% \noindent \textbf{Implications for Platform Operators and Policy Makers.}
% The transition from cheap signals to cryptographic indices has been executed at the infrastructure level in adjacent domains.
% For example, Sender Policy Framework (SPF)/Domain Keys Identified Mail (DKIM) for email authentication~\cite{}, OpenID Connect (OIDC)-backed trusted publishing for package provenance~\cite{}, and reproducible builds for artifact verification~\cite{}. 
% In these precedents, the index was enforced as a platform baseline rather than an individual option.
% % removing the two-tier dynamic in which sophisticated practitioners use cryptographic signals and unsophisticated ones do not. 
% In the software supply chain, the equivalent would be platform-level enforcement of signed provenance for packages above a usage threshold.
% With this enforcement, the use of these signals is not a recommendation that practitioners can ignore, yet it is a precondition for publication or inclusion in a registry.
% \appendix

\subsection{To Conclude: Where the Ecosystem Stands}
We opened by asking whether practitioners read the source code or rely on signals.
Both corpora confirm the latter, and those signals are collapsing as AI erodes the cost of producing them honestly versus faking them.
The software supply chain is drifting toward the condition Akerlof predicted for any market where quality signals lose their informational content: buyers cannot distinguish good from bad, good sellers cannot credibly signal their quality, and the market drifts toward lemons~\cite{akerlof1970market}.
% We opened with Akerlof's prediction for any market where quality signals lose their informational content: buyers cannot distinguish good from bad, good sellers cannot credibly signal their quality, and the market drifts toward lemons~\cite{akerlof1970market}.
% Both corpora document the software supply chain moving in this direction, not as a future risk but as a present trajectory.
This trajectory is visible in the industrialized economics of signal manipulation, the structural indistinguishability of honest and dishonest signals, and the gap between what the literature advises and what practitioners adopt.
Signaling theory offers a clear exit: shift trust away from cheap, gameable signals toward signals that AI cannot cheaply generate, i.e., cryptographic indices whose production is causally bound to the quality they certify~\cite{spence1973,searcy2005evolution}, institutional attestation~\cite{newman2022sigstore}, and accountable identity~\cite{lamb2021reproducible}.
The ecosystem can either take that exit through the institutional enforcement that Olson~\cite{olson1965logic} predicts is necessary, or continue accumulating verification overhead around gameable signals.
This response would determine whether open-source reuse retains the low-cost trust advantage that made it the lubricant of modern software development~\cite{arrow1974limits}.
% This study provides the cross-signal evidence base from which that determination will need to be made.

\begin{acks}
% TODO: For the submission, don't include acknowledgments since they would most likely deanonymize you.
This work was supported and funded by \textit{[retracted for double-blind]}. 
Any opinions, findings, and conclusions expressed in this work are those of the author(s) and do not necessarily reflect the views of \textit{[retracted for double-blind]}.
\end{acks}
 % TODO: replace with your brilliant paper!

\bibliographystyle{ACM-Reference-Format}
\bibliography{ccs-sample,short}

\end{document}